\documentclass[12pt,aps,prb,preprint,groupedaddress]{revtex4}   

\usepackage{amsmath}    
\usepackage{amsfonts}  
\usepackage{amssymb}
\usepackage{graphicx}   

\newcommand{\sDirac}[3]{\ensuremath{\langle #1|#2|#3\rangle}}
\newcommand{\dDirac}[4]{\ensuremath{\langle #1|#2 \, #3|#4\rangle}}
\newcommand{\dEnot}{\ensuremath{\Delta E_o}}
\newcommand{\commutator}[2]{\ensuremath{\left[#1,#2\right]}}
\newcommand{\op}[1]{\ensuremath{\mathbf{#1}}}
\newcommand{\efield}{\varepsilon}

\begin{document}

\title{The static electric polarizability of a particle bound by a finite potential well}
\author{M.A. Maize\cite{email} and M.A. Antonacci}
\affiliation{Department of Physics, Saint Vincent College, 300 Fraser Purchase Road, Latrobe, Pennsylvania, 15650}
\date{\today}

\begin{abstract}
In this paper we derive an expression for the static electric polarizability of a particle bound by a finite potential well without the explicit use of the continuum states in our calculations.  This will be accomplished by employing the elegant Dalgarno-Lewis perturbative technique.
\end{abstract}

\maketitle

\section{Introduction}
\label{sec:Intro}
The study of the finite potential well model and its applications has played a significant part in our fundamental education for a long time.  The model provides us with the educational experience of studying a spectrum of wavefunctions which includes both bound and continuum (unbound) states.  Second, by taking the appropriate limits, the finite potential model connects with the models of the infinite potential well and the attractive Dirac-delta potential.  This approach helps us in relating the features of the different models, and provides us with a useful approach for checking our results.  Third, the bound states of the well produce the probability of the existence of the particle in the classically forbidden region (region where the total energy of the system is less than its potential energy).  This feature helps with clarifying the differences between classical mechanics and quantum mechanics and we will study in our work the significance of the wavefunction in the classically forbidden region in determining the electric polarizability.

The applications of the finite potential well are numerous and exist in many branches of physics.  A well known example of a basic application is the modeling of the nucleon-nucleon interaction by a finite potential well.  Another example is the modeling of the scattering of low energy electrons from atoms by the scattering of a beam of particles by a system bound by a potential well (Ramsauer Effect).\cite{one}

In our work we provide an example of the effectiveness and the simplicity of the method of Dalgarno and Lewis in determining the energy shift of a quantum level in second order perturbation theory.\cite{two}  The energy shift will be used to calculate the electric polarizability due to the interaction between a static electric field and a charged particle which is bound by a finite potential well.  As is known, the conventional method used to calculate the energy shift in the second order involves an infinite sum or an integral that contains all possible states allowed by the transition.\cite{three} Some of these states, for example, scattering (unbound) states, can be very difficult or impossible to obtain in a large number of problems.  The unperturbed state will be the only part of the system which we need to know to calculate the exact energy of that particular state to the second order when we apply the technique introduced by the Dalgarno and Lewis method.  Following our approach helps in avoiding unnecessary approximations and mathematical difficulties.\cite{three, four}This in turn allows us to pay more attention to the physics of the problem and to make our work beneficial to both advanced undergraduate and first-year graduate students.  Second, some of the results which we will obtain for the electric polarizability can not be reached if the conventional method is used.  Third, we provide one more illustration of how simple methods can aid in our learning of perturbation theory.  Fourth, in addition to our previous work on both static\cite{three, four} and dynamic\cite{five, six} electric polarizability, we are providing one more example to demonstrate that the knowledge of the continuum is not always required to calculate the energy shift of a quantum level to the second order.
	
In the next section, we give a summary of the method of Dalgarno and Lewis.  We follow by introducing the necessary models, deriving the expression for the static electric polarizability and analyzing our results.  In Section~\ref{sec:Conc}, we give some concluding remarks.

\section{Method of Dalgarno and Lewis}
\label{sec:Method}

The method of Dalgarno and Lewis is based on replacing the conventional algorithm used to calculate the energy shift in second order perturbation theory with the solution of an inhomogeneous differential equation.  We give in this section a summary of the method.  References 2 and 7, Schiff\cite{eight} and Merzbacher\cite{nine} can be checked for more details.  The ground state energy shift to the second order \dEnot\ is given by

\begin{equation}
\Delta E_o=\sum^{\infty}_{n=1}\frac{\sDirac{\psi_o}{H^{\prime}}{\psi_n}\sDirac{\psi_n}{H^{\prime}}{\psi_o}}{E_o-E_n} .
\label{eq:1}
\end{equation}
$\psi_o$ is the ground state and is occupied by a charged particle as we assume in our problem.  $E_o$ is the ground state energy.  The functions $\psi_n$'s represent all states allowed by the transition due to the interaction $H^{\prime}$.  The summation in Eq.~(\ref{eq:1}) excludes the state $\psi_o$.

The first step in eliminating the infinite sum (Eq.~(\ref{eq:1})) is to find an operator $\op{F}$ which satisfies the following relationship:

\begin{equation}
\frac{\sDirac{\psi_n}{H^{\prime}}{\psi_o}}{E_o-E_n}=\sDirac{\psi_n}{\op{F}}{\psi_o}.
\label{eq:2}
\end{equation}

With the use of the completeness of all the states $\psi_n$'s and simple manipulations, \dEnot\ can be written as

\begin{equation}
\dEnot = \dDirac{\psi_o}{H^{\prime}}{\op{F}}{\psi_o}-\sDirac{\psi_o}{H^{\prime}}{\psi_o}\sDirac{\psi_o}{\op{F}}{\psi_o}.
\label{eq:3}
\end{equation}

Now, with the use of the following property:

\begin{equation}
\sDirac{\psi_n}{\commutator{\op{F}}{H_{o}}}{\psi_o}=(E_o-E_n)\sDirac{\psi_n}{\op{F}}{\psi_o},
\label{eq:4}
\end{equation}

Eq.~(\ref{eq:2}), can be written as\cite{eight}

\begin{equation}
(E_o-H_o)\phi=H^{\prime}\psi_o-\sDirac{\psi_o}{H^{\prime}}{\psi_o}\psi_o,
\label{eq:5}
\end{equation}

where

\begin{equation}
H_o\psi_o=E_o\psi_o,
\label{eq:6}
\end{equation}

and

\begin{equation}
\phi=\op{F}\psi_o.
\label{eq:7}
\end{equation}

It should be noted that $\phi$ (Eq.~(\ref{eq:5})) can contain an arbitrary function $\chi$ with $\chi$ satisfying the relationship $(E_o-H_o)\chi=0$.  The final expression of $\phi$ is taken to be orthogonal to $\psi_o$\cite{two, seven, eight, nine} and with this choice Eq.~(\ref{eq:3}) can be written as

\begin{equation}
\dEnot=\sDirac{\psi_o}{H^{\prime}}{\phi}.
\label{eq:8}
\end{equation}

To obtain \dEnot\ using the method of Dalgarno and Lewis, we first find $\phi$ with Eq.~(\ref{eq:5}) being our starting point.  The second and final step is to use $\phi$ in Eq.~(\ref{eq:8}) to obtain \dEnot.  With these two steps, the only stationary state of the system we use in our calculation is $\psi_o$ and we completely avoid the infinite summation in Eq.~(\ref{eq:1}).

Before closing this section, we point to an important application  of the Dalgarno and Lewis method.  In our previous work of solving for the electric polarizability in the case of the delta potential,\cite{three, four} $\chi$ has been equal to zero.  In the current work, and as we will find out, $\chi$ has to exist to obtain the correct polarizability.

\section{The Electric Polarizability}
\label{sec:EP}

As stated before, the model we are presenting in this article is of a particle bound by a finite potential well.  The potential energy $V(x)$ is given by:

\begin{equation}
 V(x)= \left\{ \begin{array}{ll}
							-V_o  &  \mbox{\ \ \ \ $(-a < x < a)$, }  \\
							0     &  \mbox{\ \ \ \      (otherwise),}
							\end{array}
			\right. 
\label{eq:9}
\end{equation}
where $2a$ is the width of the well.  The one-dimensional Schr\"{o}dinger equation with the given $V(x)$ produces a spectrum of bound states in two classes: even parity states and odd parity states in addition to a continuum of unbound states.\cite{ten}  The unperturbed state $\psi_o$ in our problem is the lowest energy even parity state for a given $V_o$ and $a$.  $\psi_o$ is then given by\cite{one}

\begin{equation*}
\hspace{156pt}\psi_o(x < -a) = N \cos K_{o}a e^{k_{o}(x+a)},\hspace*{115pt}	(10.a)
\label{eq:10a}
\end{equation*}
and
\begin{equation*}
\hspace{153pt} \psi_o(-a < x < a) = N \cos K_{o}x,\hspace*{132pt}	(10.b)
\label{eq:10b}
\end{equation*}
and
\begin{equation*}
\hspace{156pt} \psi_o(x > a) = N \cos K_{o}a e^{-k_{o}(x-a)},\hspace*{115pt}	(10.c)
\label{eq:10c}
\end{equation*}
\addtocounter{equation}{1} 
where $E_o$ is the eigenenergy of the state $\psi_o$ and is negative, $K^{2}_{o}=\frac{2m}{\hbar^2}(V_o - \left|E_o\right|)$, $k_{o}^{2}=\frac{2m\left|E_o\right|}{\hbar^2}$ and $K_{o}a < \frac{\pi}{2}$.  The normalization constant N is given by

\begin{equation}
N = \frac{1}{\left[a+\frac{\sin{K_{o}a}\cos{K_{o}a}}{K_o}+\frac{\cos^2{K_{o}a}}{k_o}\right]^\frac{1}{2}}.
\label{eq:11}
\end{equation}
The determination of the eigenvalues of the well bound states is conventionally obtained via finding the roots of certain transcendental equations.\cite{eleven}  For a given $V_o$ and $a$, the transcendental equations needed to determine $E_o$ are given by

\begin{equation}
\gamma_{o}\tan{\gamma_o}=\beta_{o},
\label{eq:12}
\end{equation}
and
\begin{equation}
\gamma_{o}^{2}+\beta_{o}^{2}=R^{2},
\label{eq:13}
\end{equation}
where $\gamma_o=K_{o}a$, $\beta_{o}=k_{o}a$ and $R^2=\frac{2ma^{2}V_o}{\hbar^2}$.  

The conventional method as it appears in many textbooks,\cite{one, eight, nine, ten} does not provide the simplest procedure for determining the bound states' eigenvalues.  Due to this a variety of useful approximations had been developed to replace the conventional method.\cite{ eleven, twelve, thirteen}  In Ref. 13, the authors used their approach of simplifying Eqs.~(\ref{eq:12}) and~(\ref{eq:13}) to approximate the finite potential well by a wider infinite potential well.  The width of the infinite potential well is equal to the width of the finite potential well multiplied by the factor $(1+\frac{1}{R})$.  The resulting approximate energy eigenvalues range in their accuracy with the ground state energy being the most accurate (less than 1\% error) in the range of $R$ close to or larger than four.~\cite{thirteen}  The results of Ref. 13, which we will get back to, will be useful in our work when we test our expression of the electric polarizability in the limiting case of large $R$.

To determine the energy shift $\dEnot$, we study the interaction between a particle of charge $q$ that is initially in the state $\psi_o$ and an external static electric field $\efield$ which is applied in the vicinity of the particle.  Once we find $\dEnot$, the electric polarizability $\alpha$ can be determined from the relationship $\dEnot=-\frac{1}{2}\alpha\efield^2$.  Now, to simplify the geometry of the problem without any loss of generality, we take the electric field $\efield$ to be parallel to the $x$ axis.  The electric dipole Hamiltonian which represents the interaction of the electric field and the charge is then given by

\begin{equation}
H^{\prime}=-q\efield x.
\label{eq:14}
\end{equation}
At this point we are in a position to calculate $\dEnot$ and the polarizability.  With $H_o=\frac{-\hbar^{2}}{2m}\frac{\partial^2}{\partial x^2}+V(x)$ where m is the mass of the particle and with the aid of Eqs. (10.a) and (10.c), we can write Eq.~(\ref{eq:5}) for the region $x>\left|a\right|$ as follows

\begin{equation}
\left[\frac{-\hbar^{2} k_{o}^{2}}{2m} + \frac{\hbar^2}{2m} \left(\frac{\partial^2}{\partial x^2}\right)\right] \phi \left(x<-a\right) = -q \efield x N \cos K_{o}a e^{k_{o}\left(x+a\right)},
\label{eq:15}
\end{equation}
and
\begin{equation}
\left[\frac{-\hbar^{2} k_{o}^{2}}{2m} + \frac{\hbar^2}{2m} \left(\frac{\partial^2}{\partial x^2}\right)\right] \phi \left(x>a\right) = -q \efield x N \cos K_{o}a e^{-k_{o}\left(x-a\right)},
\label{eq:16}
\end{equation}
where $\phi\left(x<-a\right)$ and $\phi\left(x>a\right)$ refer to the expressions of $\phi$ in the regions $x<-a$ and $x>a$ respectively.  The second term on the right-hand side of Eq.~(\ref{eq:5}) vanishes because $H^\prime$ has odd parity and $\psi_o$ has a definite parity.

Equations~(\ref{eq:15}) and~(\ref{eq:16}) are very simple and can be solved essentially by inspection.  The expressions for $\phi\left(x<-a\right)$ and $\phi\left(x>a\right)$ are given by
\begin{equation}
\phi\left(x<-a\right) = \left[-\left(\frac{mq\efield}{2\hbar^{2}k_o}\right)x^{2}+\left(\frac{mq\efield}{2\hbar^{2}k_{o}^{2}}\right) x\right] N \cos K_{o}a  e^{k_{o}\left(x+a\right)},
\label{eq:17}
\end{equation}
and
\begin{equation}
\phi\left(x>a\right) = \left[\left(\frac{mq\efield}{2\hbar^{2}k_o}\right)x^{2}+\left(\frac{mq\efield}{2\hbar^{2}k_{o}^{2}}\right) x\right] N \cos K_{o}a e^{-k_{o}\left(x-a\right)}.
\label{eq:18}
\end{equation}
Substituting the expressions of $\phi\left(x>a\right)$, $\psi\left(x>a\right)$, $\phi\left(x<-a\right)$ and $\psi\left(x<-a\right)$ in Eq.~(\ref{eq:8}), and performing the necessary integration, we obtain the energy shift corresponding to $\psi_o$ in the region $x>\left|a\right|$.  The electric polarizability, $\alpha_1$, corresponding to this value of the energy shift can be written as
\begin{equation}
\alpha_1 = \left(\frac{mq^2N^2}{\hbar^2}\right)\cos^2K_o a \left[\frac{a^3}{k_{o}^{2}} + \frac{5a^2}{2k_{o}^{3}} + \frac{5a}{2k_{o}^{4}} + \frac{5}{4k_{o}^{5}}\right].
\label{eq:19}
\end{equation}
By taking the infinite potential well limit $\left(K_o a \rightarrow \frac{\pi}{2}\right)$, $\alpha_1$ becomes zero.  Now by taking the limits of $2a$ approaching $0$, with $V_o$ approaching infinity while keeping $a V_o$ constant, we can obtain $\alpha_1$ for the case of the attractive delta.  For the given limits, $N^2\cos^2 K_o a$ will become $k_o$ and $\alpha_1$ is then given by $\frac{5}{4}\left[\frac{\left(mq^{2}\right)}{\left(\hbar^{2}k_{o}^{4}\right)}\right]$ which is the correct result for the polarizability in the case of the attractive delta.~\cite{three}

To evaluate the contribution to \dEnot\ from the region $-a<x<a$, we first solve Eq.~(\ref{eq:5}) for $\phi$ in this region.  Eq.~(\ref{eq:5}) for this region is given by
\begin{equation}
\left[-\left(\frac{\hbar^{2}k_{o}^{2}}{2m}\right)+\frac{\hbar^{2}}{2m}\frac{\partial^{2}}{\partial x^2}+V_o\right]\phi_t\left(-a<x<a\right)=-q\efield xN\cos K_o x,
\label{eq:20}
\end{equation}
where $\phi_t$ is a trial function which will satisfy Eq.~(\ref{eq:20}).  $\phi_t$ can be found by a simple process of trial and error and it is given by
\begin{equation}
\phi_t\left(-a<x<a\right)=\left(\frac{-mq\efield N}{2 \hbar^2 K_o}\right)x^2 \sin K_o x +\left(\frac{-mq\efield N}{2 \hbar^2 K_{o}^{2}}\right)x \cos K_o x.
\label{eq:21}
\end{equation}
The expressions of $\phi_t\left(-a<x<a\right)$ and $\psi_o\left(-a<x<a\right)$ are then used in Eq.~(\ref{eq:8}) to find the energy shift corresponding to $\psi_o$ in the region $-a<x<a$.  The electric polarizability $(\alpha_2)_t$ corresponding to this energy shift is given by
\begin{equation}
\left(\alpha_2\right)_t=\left(\frac{mq^2 N^2}{\hbar^2}\right)\left[\frac{-a^3}{3K_{o}^{2}}+\frac{a^3}{2K_{o}^{2}}\cos 2K_o a-\frac{5a^2}{4K_{o}^{3}}\sin 2K_o a-\frac{5a}{4 K_{o}^{4}} \cos 2 K_o a + \frac{5}{8 K_{o}^{5}} \sin 2 K_o a \right].
\label{eq:22}
\end{equation}
By taking the limit corresponding to the attractive delta, $\left(\alpha_2\right)_t$ becomes zero.  By taking the infinite potential well limit $\left(K_o a \rightarrow \frac{\pi}{2}\right)$, $\left(\alpha_2\right)_t=-0.1324176\left(\frac{mq^2 a^4}{\hbar^2}\right)$.  At this point the electric polarizability $\alpha$ is equal to $\left[\alpha_1+\left(\alpha_2\right)_t\right]$.  $\alpha_1$ gives zero for the case of the infinite potential well as expected since $\psi_o$ for the region $x>\left|a\right|$ represents the system outside the well.  The contribution to $\alpha$ for the case of the infinite potential well then comes from $\left(\alpha_2\right)_t$ and this means that $\alpha$ is negative.  Checking the expression of \dEnot\ (Eq.~(\ref{eq:1})) and the definition of $\alpha$ $\left(\dEnot=-\frac{1}{2}\alpha\efield^2\right)$, we conclude that $\alpha$ should be positive.  So, even though the expression of $\phi_t\left(-a<x<a\right)$ satisfies Eq.~(\ref{eq:20}), it can not be a complete expression for $\phi$ in the region $-a<x<a$.

As we learned from the previous section, $\phi$ can contain an arbitrary function $\chi$ with the condition $\left(E_o-H_o\right)\chi=0$.  We take $\chi$ to be equal to $\left(\frac{-mq\efield N}{2 \hbar^2 K_o}\right)C\sin K_o x$ with the coefficient $C$ being $x$ independent.  The reasons for making this choice for $\chi$ are: (1) $\left(\phi_t+\chi\right)$ is expected to be orthogonal to $\psi_o$;\cite{two, seven, eight, nine} (2) $\chi$ should provide positive contribution to $\alpha$;  (3) The structure of $\chi$ should be consistent with the structure of $\phi_t$.

To determine the coefficient $C$, we follow two simple steps.  The first step is to calculate $\alpha$ for the infinite potential well with the use of the Dalgarno-Lewis technique and the conventional method.  The comparison of the two results will then produce $C_{inf.pot.}$ ($C$ in the case of the infinite well).  In applying the Dalgarno-Lewis technique we calculate $\alpha$ by replacing $\phi$ by $\phi_t + \chi$ in Eq.~(\ref{eq:8}), setting $K_o=\frac{\pi}{2a}$, and performing the necessary integration.  In using the conventional method, we have to use Eq.~(\ref{eq:1}), however the equation contains an infinite sum.  The infinite sum causes no problem, since the numerator inside the sum decreases rapidly with increasing $n$, and the denominator is proportional to $(n^2 - 1)$.  A simple calculation using the spectrum of the infinite well will show that the transition $n=1$ to $n=2$ is far superior in magnitude to the next few terms.  Using only the transition $n=1$ to $n=2$, and comparing the results of both the conventional and the Dalgarno-Lewis method, we find that the choice $C_{inf.pot.}=-a^2$ produces an almost perfect agreement.  The one-term conventional method produces a polarizability of $0.0701371\left(\frac{mq^2 a^4}{\hbar^2}\right)$ while substituting by $C_{inf.pot.}=-a^2$ in the other expression of the polarizability (Dalgarno-Lewis) produces $\alpha_{inf.pot.}$ which is given by:
\begin{equation}
\alpha_{inf. pot.}=0.0702247\left(\frac{mq^2 a^4}{\hbar^2}\right).
\label{eq:23}
\end{equation}
The second step is to use the result $C_{inf.pot.}$ to help us predict $C$.  Our prediction should produce the right result for the infinite potential well, satisfy the limiting case of the attractive delta potential and be consistent with the structure of $\phi_t$.  The choice of $C=\frac{-\left(\frac{\pi}{2}\right)^2}{K_{o}^{2}}$ clearly fulfills the first and third conditions and its fulfillment of the second condition will be demonstrated.  In addition testing our choice of $C$ away from  the limit $K_o a \rightarrow \frac{\pi}{2}$ is warranted.

At this point and with our choice of $C$, $\phi\left(-a<x<a\right)$ and the corresponding electric polarizability $\alpha_2$ are given by
\begin{equation}
\phi\left(-a<x<a\right)=\left(\frac{-mq\efield N}{2 \hbar^2 K_o}\right)\left[x^2\sin K_o x+\frac{x}{K_o}\cos K_o x-\frac{\left(\frac{\pi}{2}\right)^2}{K_{o}^{2}}\sin K_o x\right],
\label{eq:24}
\end{equation}
and
\begin{equation}
\alpha_2=\left(\frac{mq^2 N^2}{\hbar^2}\right)\left[\frac{-a^3}{3K_{o}^{2}}+f_1\left(K_o,a\right)\cos 2K_o a + f_2\left(K_o,a\right)\sin 2 K_o a\right],
\label{eq:25}
\end{equation}
where
\begin{equation}
f_1\left(K_o,0\right)=\frac{a^3}{2K_{o}^{2}}-\frac{5a}{4K_{o}^{4}}-\frac{\left(\frac{\pi}{2}\right)^2 a}{2 K_{o}^{4}},
\label{eq:26}
\end{equation}
and
\begin{equation}
f_2\left(K_o,0\right)=\frac{-5a^2}{4K_{o}^{3}}+\frac{5}{8K_{o}^{5}}+\frac{\left(\frac{\pi}{2}\right)^2}{4 K_{o}^{5}}.
\label{eq:27}
\end{equation}
By taking the limits corresponding to the attractive delta, $\alpha_2$ becomes zero.  We take the electric polarizability $\alpha$ to be equal to $\alpha_1+\alpha_2$.

To test our choice of $C$ away from the infinite well limit, we use the results of Ref. (13) to write an approximate expression for the polarizability ($\alpha_{apr}$) and use this expression to compare with our expression of $\alpha$.  According to the work of Ref. (13), a finite potential well of width $2a$ can be approximated by a wider infinite potential well of with $2b$ with $b=\left(1+\frac{1}{R}\right)a$.  The approximation produces small error in the eigenvalues and the probability densities of the eigenfunctions belonging to the lowest energy states for $R$ close to or larger than four.\cite{thirteen}  For example, in comparing the approximated and the exact energy eigenvalue for the lowest two energy states at $R=4$, it is 0.69\% for the lowest energy even parity state, and 3.15\% for the lowest energy odd parity state.  The approximation also produces very small error for the probability densities of the two states as a function of $x$ (Fig. 3 of Ref. (13)).  These results are significant to our intended comparison due to two main reasons.  First, the lowest energy state is the only state of the system which we use in our calculation of the polarizability. Second, the transition $n=1$ to $n=2$ produces almost all the contribution to the electric polarizability in the case of the infinite potential well.

To present our comparison, we first write $\alpha_{apr}$ by replacing $a$ in Eq.~(\ref{eq:23}) by b.  $\alpha_{apr}$ is then given by
\begin{equation}
\alpha_{apr}=0.0702247 \left(\frac{mq^2}{\hbar^2}\right)\left(1+\frac{1}{R}\right)^{4}a^4
\label{eq:28}
\end{equation}
Now we set $g=\left(\frac{mq^2}{\hbar^2}\right)a^4$ and divide $\alpha_1$, $\alpha_2$ and $\alpha_{apr}$ by $g$ to get the dimensionless polarizabilities $\alpha_{1}^{\prime}$, $\alpha_{2}^{\prime}$ and $\alpha_{apr}^{\prime}$.  $\alpha_{1}^{\prime}$ and $\alpha_{2}^{\prime}$ will be given by
\begin{equation}
\alpha_{1}^{\prime}=N^{\prime2} \cos^2 \gamma_o \left[\frac{1}{\beta_{o}^{2}} + \frac{5}{2\beta_{o}^{3}}+\frac{5}{2\beta_{o}^{4}}+\frac{5}{4\beta_{o}^{5}}\right],
\label{eq:29}
\end{equation}
and
\begin{equation}
\alpha_{2}^{\prime}=N^{\prime2}\left[\frac{-1}{3\gamma_{o}^{2}}+f^{\prime}_{1}\left(\gamma_o\right)\cos 2\gamma_o+f^{\prime}_{2}\left(\gamma_o\right)\sin2\gamma_o\right],
\label{eq:30}
\end{equation}
where
\begin{equation}
f_{1}^{\prime}\left(\gamma_o\right)=\frac{1}{2\gamma_{0}^{2}}-\frac{5}{4\gamma_{o}^{4}}-\frac{\left(\frac{\pi}{2}\right)^2}{2\gamma_{o}^{4}},
\label{eq:31}
\end{equation}
and
\begin{equation}
f_{2}^{\prime}\left(\gamma_o\right)=\frac{-5}{4\gamma_{o}^{3}}+\frac{5}{8\gamma_{o}^{5}}+\frac{\left(\frac{\pi}{2}\right)^2}{4\gamma_{o}^{5}}.
\label{eq:32}
\end{equation}
Accordingly $N^{\prime2}$ is written as
\begin{equation}
N^{\prime2}=\frac{1}{\left[1+\frac{\sin\gamma_o\cos\gamma_o}{\gamma_o}+\frac{\cos^2\gamma_o}{\beta_o}\right]},
\label{eq:33}
\end{equation}
and $\alpha^{\prime}=\alpha_{1}^{\prime}+\alpha_{2}^{\prime}$.  In Table~\ref{Table 1}, we present the numerical values for $\alpha_{1}^{\prime}$, $\alpha_{2}^{\prime}$, $\alpha^{\prime}$ and $\alpha_{apr}^{\prime}$ for $R$ values ranging from around four to $R$ becoming very large.  This is the range of $R$ where the results for $\alpha_{apr}$ are reliable according to the previous discussion.  Also, in this range of $R$, $\alpha_{2}^{\prime}$ is close or approximately equal to $\alpha^{\prime}$ and this provides a crucial testing to our choice of $C$.  Now in comparing the numerical results for $\alpha^{\prime}$ and $\alpha_{apr}^{\prime}$ we find that we have an excellent agreement.  Of course, we did not expect a perfect agreement because $\alpha_{apr}^{\prime}$ is based on an excellent approximation rather than exact calculations.  To see how our choice of $C$ contributes to the agreement between the numerical results of $\alpha^{\prime}$ and $\alpha_{apr}^{\prime}$, we define $T$ as $\left(\frac{\alpha_{2}^{\prime}-\left(\alpha_{2}^{\prime} \right)_{t}}{\alpha_{2}^{\prime}}\right)$, with $\left(\alpha_{2}^{\prime}\right)_t=\frac{\left(\alpha_2\right)_{t}}{g}$.  $T$ can be obtained by calculating the term $\left[\frac{-\left(\frac{\pi}{2}\right)^2}{2 \gamma_{o}^{4}}\cos 2\gamma_o + \frac{\left(\frac{\pi}{2}\right)^2}{4\gamma_{o}^{5}}\sin 2 \gamma_o \right]$ and dividing the outcome by the bracket on the R.H.S. of Eq.~(\ref{eq:30}) for a given $\gamma_o$.  For $\gamma_o=0.39\pi$, we obtain $T=2.52$, and with $\gamma_o=0.47\pi$, $T=2.84$.  This demonstrates that the choice of $C$ is crucial to the obtained value of $\alpha_{2}^{\prime} \left(\alpha_{2}\right)$.  With this result, the observed dominance of $\alpha_{2}^{\prime}$ over $\alpha_{1}^{\prime}$ in determining $\alpha^\prime$ (Table~\ref{Table 1}) and the agreement of the numerical result of $\alpha^{\prime}$ and $\alpha_{apr}^{\prime}$, we have verified the validity of our choice of $C$.  The result of $\alpha=\alpha_1 + \alpha_2$ is then our final expression for the electric polarizability of the finite well.

Referring to Table~\ref{Table 2}, we present our results for $\alpha_{1}^{\prime}$, $\alpha_{2}^{\prime}$ and $\alpha^{\prime}$ in the region of $\gamma_{o} < 0.2\pi$.  In this region $\alpha_{1}^{\prime}$ which has been obtained with the aid of $\psi_{o}\left( x > \left|a\right|\right)$, is approximately equal to $\alpha^{\prime}$.  In addition, we already verified that the delta-potential polarizability is obtained from $\alpha_{1}$ when the appropriate limits are taken.  These results demonstrate the superiority of the bound state in the classically forbidden region in determining the electric polarizability in the region of smaller $\gamma_{o}$.

Considering the dependence of $\alpha^{\prime}$ on $R$, our results in both tables in addition to the results which are not shown here demonstrate that as $R$ decreases, $\alpha^{\prime}$ increases.  With our definition of $\alpha^{\prime}$ ($\alpha^{\prime}=\frac{\alpha}{g}$), we fixed $a$ and this means that a smaller $R$ leads to a lower binding energy, i.e. lower $\left| E_o \right|$.  Since the polarizability is inversely proportional to the energy gap between the ground state and the higher energy states (Eq.~(\ref{eq:1}) and the basic definition of polarizability), we then expect $\alpha^{\prime}$ to be increasing while $R$ is decreasing.

\section{Conclusion}
\label{sec:Conc}

In this paper we have derived  an expression for the static electric polarizability of a particle bound by a finite potential well.  The only stationary state of the system which we have used in our calculation is the unperturbed state of the particle.  This result and the studies appearing in our previous work,\cite{three, four} teach us that the expression of the continuum is not required for every calculation of the electric polarizability.  It is the elegance of the Dalgarno-Lewis method which has allowed us to avoid any complications arising from dealing with the continuum and to have only simple functions (sines and cosines) appearing in our expressions.  The simplicity of the method and expressions helps the reader in checking every step of our work.  It aids in clarifying the physics of the problem and certainly encourages physics students to learn about perturbation theory via the application of the Dalgarno-Lewis method.

In deriving the expression for the static electric polarizability in the case of the delta-function,\cite{three, four} the solution of Eq.~(\ref{eq:5}) gave us the final expression for $\phi$.  In the case of the finite potential well a function $\chi$ has to be added to the solution of Eq.~(\ref{eq:5}).  In determining $\chi$, we used basic physics principles, the model of the infinite potential well, and the approximation of Ref~(13).  This process in turn has provided us with some valuable lessons.  First, the use of the Dalgarno-Lewis method in calculating the electric polarizability can be extended beyond problems which commonly appear in literature.  Among such problems are the cases of the single delta, the infinite potential well, and the hydrogen atom in its ground state.  Second, we avoided the infinite sum of the conventional method, but we still used the first term of the sum in the case of the infinite potential well to determine $\chi$.  In this process, we continue to depend only on bound states in the simplest form to obtain our result.  Third, it is the realization that the approximation of Ref.~(13) has practical applications in solving problems related to the finite potential well.

The starting point in our derivation has been the solution of Eq.~(\ref{eq:5}).  This allowed us to write the contribution of $\alpha$ in terms of $\alpha_1$ and $\alpha_2$.  This separation has been helpful in clarifying how the bound state in the classically forbidden region contributes to the electric polarizability.  In addition, we can specify that $\alpha_1$ is what produces the limit for the delta potential while $\alpha_2$ produces the limit for the infinite potential well.  Such separation would be impossible in the case of the conventional method since the determination of $\alpha$ depends on the calculation of the term $\left|\sDirac{\psi_n}{H^{\prime}}{\psi_o}\right|^2$ (Eq.~(\ref{eq:1})).

\begin{table}[p]
	\centering
									\begin{tabular*}{6.5in}{@{\extracolsep{\fill}}c@{\extracolsep{\fill}}c@{\extracolsep{\fill}}c@{\extracolsep{\fill}}c@{\extracolsep{\fill}}c@{\extracolsep{\fill}}c@{\extracolsep{\fill}}c}
		\hline							
		$\gamma_o$	&$\beta_o$	&		$R$	  &	$\alpha_{1}^{\prime}$	&	$\alpha_{2}^{\prime}$	&	$\alpha^{\prime}$& $\alpha_{apr}^{\prime}$ \\
		\hline \hline
		0.39$\pi$		&	3.403183	& 3.617018	&	0.015178						&	0.173148						&	0.188326			 	&	0.186438 \\
		0.41$\pi$		& 4.433507	& 4.616825	& 0.005510						& 0.147482						& 0.152993			 	&	0.153844 \\
		0.43$\pi$		& 6.043511	& 6.192650	& 0.001663						&	0.125180						& 0.126843			 	&	0.127803 \\
		0.45$\pi$		&	8.925856	& 9.037118	&	0.000363						& 0.106019						&	0.106382			 	&	0.106858 \\
		0.47$\pi$		&15.620252	&15.589884	&	3.99E-5							&	0.089754						&	0.089794			 	&	0.089913 \\
		0.49$\pi$		&48.983879	&49.008061	&	4.24E-7							& 0.076129						&	0.076129			 	&	0.076134 \\
		\hline
		\end{tabular*}
	\caption{Comparison of the dimensionless polarizability $\alpha^{\prime}$ (our calculation with the Dalgarno-Lewis method) with the dimensionless polarizability $\alpha_{apr}^{\prime}$ (our calculation with the infinite potential well approximation of Ref.~(13)).}
	\label{Table 1}
\end{table}

\begin{table}[p]
	\centering
									\begin{tabular*}{6.5in}{@{\extracolsep{\fill}}c@{\extracolsep{\fill}}c@{\extracolsep{\fill}}c@{\extracolsep{\fill}}c@{\extracolsep{\fill}}c@{\extracolsep{\fill}}c}
		\hline
		$\gamma_o$	&	$\beta_o$	&	$R$	  &	$\alpha_{1}^{\prime}$	&	$\alpha_{2}^{\prime}$	&	$\alpha^{\prime}$ \\
		\hline \hline
		
		0.19$\pi$		& 0.405655	& 0.721698	&	4.93E+1						&	0.620993						&	4.99E+1	\\
		0.17$\pi$		&	0.315849	& 0.620477	&	1.31E+2						&	0.677762						&	1.32E+2  \\
		0.15$\pi$		&	0.240108	& 0.528884	&	3.87E+2						& 0.733438						& 3.88E+2 \\
		\hline
				\end{tabular*}
	\caption{Comparison of the contributions of $\alpha_{1}^{\prime}$ and $\alpha_{2}^{\prime}$ to $\alpha^{\prime}$ for $\gamma_o < 0.2\pi$.}
	\label{Table 2}
\end{table}



\end{document}